**Full Title:** When tails wag the decision: The role of distributional tails on climate impacts on decision-relevant time-scales

**Short Title:** Distributional tails in decision analysis

Gregory G. Garner[1] and Klaus Keller[2,3,4]

[1] Woodrow Wilson School of Public and International Affairs, Princeton University, Princeton, New Jersey 08544, U.S.A.

[2] Earth and Environmental Systems Institute, The Pennsylvania State University, University Park, Pennsylvania 16803, U.S.A.

[3] Department of Geosciences, The Pennsylvania State University, University Park, Pennsylvania 16803, U.S.A.

[4] Department of Engineering and Public Policy, Carnegie Mellon University, Pittsburgh, Pennsylvania 15213, U.S.A.

**Abstract**

Assessing and managing risks in a changing climate requires projections that account for decision-relevant uncertainties. These deep uncertainties are often approximated by ensembles of Earth-system model runs that sample only a subset of the known uncertainties. Here we demonstrate and quantify how this approach can cut off the tails of the distributions of projected climate variables such as sea-level rise. As a result, low-probability high-impact events that may drive risks can be under-represented. Neglecting the tails of this deep uncertainty may lead to overconfident projections and poor decisions when high reliabilities are important.

**Introduction**

Many decisions we make today depend on projections of the complex and dynamic Earth-system. For example, designing a strategy to manage coastal flooding risks depends on projections of future sea levels, storm surges, and human behavior [1,2]. Decision makers often have strong preferences for the reliable functioning of crucial systems [2]. As a result, decision analyses often need to account for the relevant uncertainties that affect these reliabilities [3].

Some risk and decision analyses sample potential future climates using an ensemble of complex Earth-system models, such as those in the Climate-Model Intercomparison Project (CMIP) [4–7]. These ensembles have provided valuable insights, but it is important to remember that these ensembles are often a collection of different best-estimates of the system [8,9], thus sampling only a subset of the known uncertainties [10–12] often due to limited computational resources, limited knowledge of the system, or scarcity of data and information [13–15]. As a result, the empirical distribution from these ensembles may be appropriate for some decision objectives (e.g., to perform well in the most likely outcome), but decision makers often care about low-probability high-impact events [16,17]. For example, designing coastal-defense infrastructure often strives for infrequent breaches on the order of one per several decades to thousands of years [18]. The decision makers here require information about the extreme tails. A distribution of best-estimates – in contrast – focuses on the mode of the distribution and may well miss decision relevant tails. Some studies have started to address this issue by post-processing these best-estimates to remove ensemble-member correlations [19], by fusing observations and models using a Bayesian modeling framework [20,21], or by adopting a framework from the rapidly growing field of robust decision making (RDM) [22–24]. However, the underlying problem of the limited number of ensemble members persists.

**Data and Methods**

We consider a simple example driver of projection uncertainty in climate change: the equilibrium climate sensitivity [25]. Equilibrium climate sensitivity is the change in the global equilibrium surface air temperature due to a doubling of the atmospheric carbon dioxide concentration relative to pre-industrial levels [26]. Climate sensitivity is a deeply uncertain quantity [27–30] (Fig 1) and an important characteristic of the climate system [31]. Current estimates of climate sensitivity are poorly characterized by a single estimate or range of best-estimates. The range of the empirical distribution of 43 complex-model estimates [25] (Fig 1 – vertical bar) gives zero probability to anything beyond the highest value sampled by the ensemble. In contrast, the estimates of the climate sensitivity distribution considered here [27–30] produce substantial probability in the tails beyond the range of the empirical distribution. Incorporating the deeply-uncertain tails of these climate sensitivity estimates into climate risk assessment can be crucial to informing decisions [32,33].

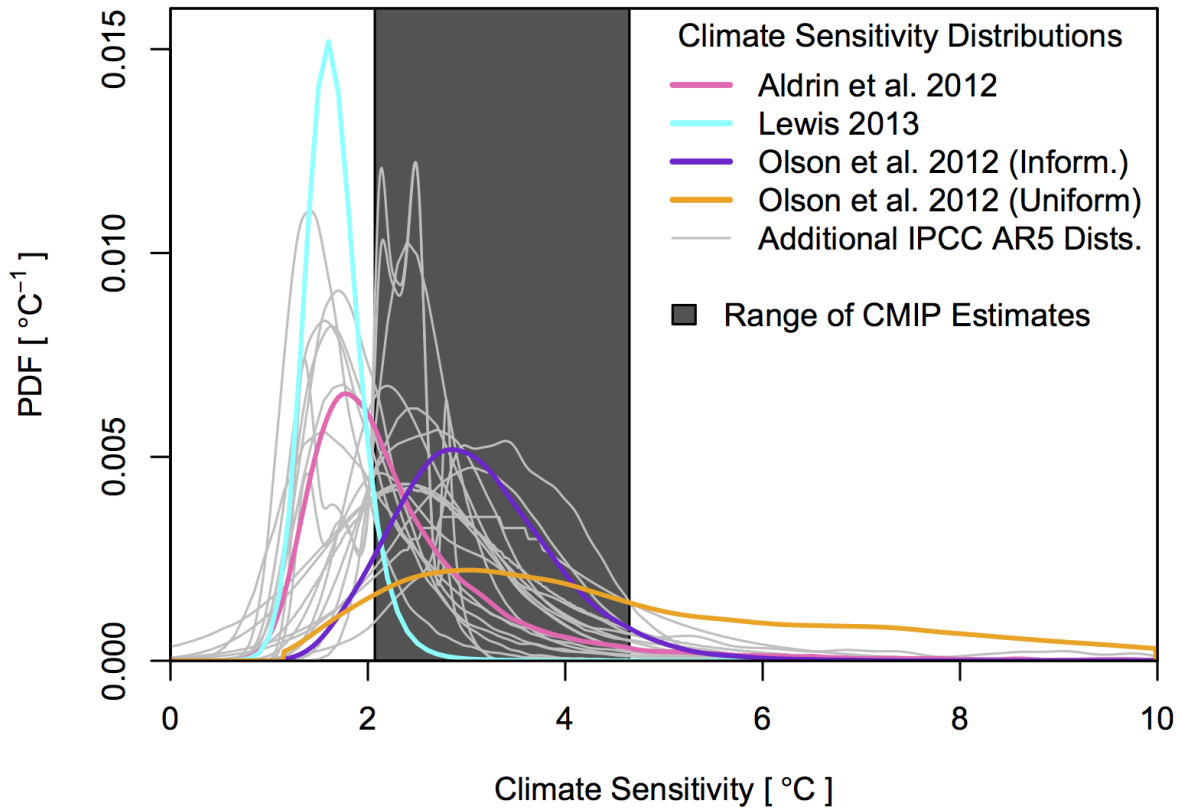

**Fig 1. Empirical probability distributions of climate sensitivity.**

The vertical black bar is the range of the 43 climate sensitivity estimates from CMIP 3 and CMIP 5 models [25]. Each of the gray lines is a climate sensitivity distribution. Four of the distributions that generally span the range of the collective 23 distributions are highlighted in color.

To demonstrate the mapping of uncertainty in climate sensitivity to projections of atmospheric temperature, we use the simplest possible model structures we could identify that can demonstrate the effects in question. Much more complex model structures are, of course, possible [34,35], but these more complex model structures would introduce considerable and arguably unnecessary computational demands and complexity in this exposition.

For one, we use the Diffusion Ocean Energy balance Climate Model (DOECLIM) [36]. DOECLIM is a one-dimension energy-balance model with a diffuse ocean model. It captures the relationship between climate sensitivity and the diffusion of heat energy into the oceans [26]. The model inputs are an exogenous time-series of atmospheric forcing on an annual time-step, a single value of climate sensitivity, and a single value of ocean diffusivity. Atmospheric temperature, sea-surface temperature, and heat fluxes are calculated endogenously at an annual time-step. Since we will be using samples from distributions of climate sensitivity, we must first calibrate the model to ensure the pair of climate sensitivity and ocean diffusivity values used are consistent with a historical temperature record [37].

We calibrate the ocean diffusivity parameter to the selected climate sensitivity by producing temperature hindcasts using historical radiative forcing provided from the Radiative Concentration Pathway (RCP) database [38]. In these hindcasts, we sample climate sensitivity and select the ocean diffusivity value that minimizes the sum of square errors between the National Aeronautics and Space Administration (NASA) Goddard Institute for Space Studies (GISS) historical temperature record [39,40] and the model hindcast [37]. More complex methods can be used to estimate the joint distributions of the two parameters [28]; however, our simplified approach provides a reasonable calibration of climate sensitivity to ocean diffusivity (Fig 2) consistent with the historical temperature record (Fig 3) and is sufficient to demonstrate our points.

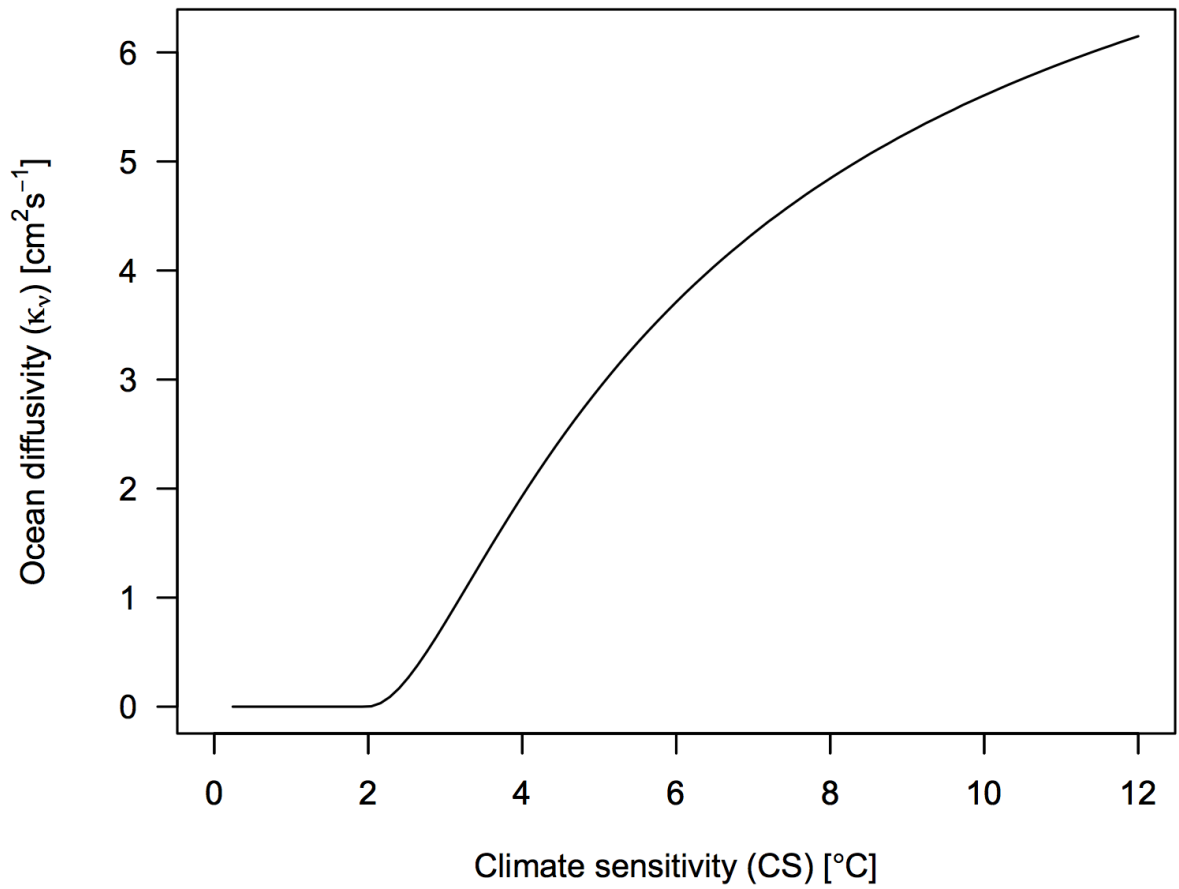

**Fig 2. DOECLIM model calibration results.**

The ocean diffusivity parameter in the DOECLIM model, which regulates the rate at which heat is vertically diffused into the ocean, is calibrated to climate sensitivity. This calibrated relationship ensures temperature hindcasts are consistent with the historical temperature record.

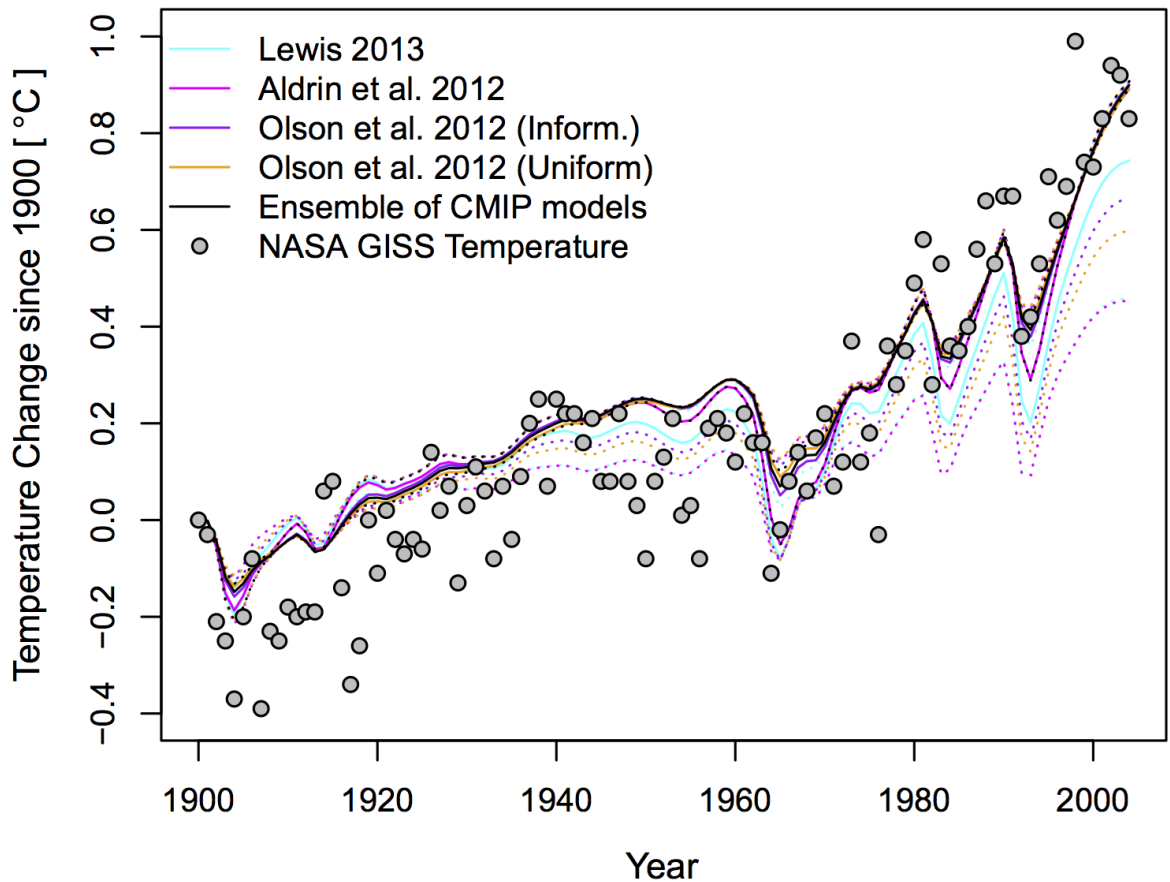

**Fig 3. DOECLIM temperature hindcasts.**

The gray dots are the National Aeronautics and Space Administration – Goddard Institute for Space Studies observed mean annual global temperature anomaly from year 1900 [39,40]. The lines represent the temperature hindcasts from the highlighted distributions and Climate Model Intercomparison Project model ensemble [25]. Solid lines are the median and dotted lines are the 0.5[th] and 99.5[th] percentile of the respective distributions.

We project atmospheric temperature with the calibrated DOECLIM model, the climate sensitivity distributions, and the equivalent radiative forcing of the RCP8.5 pathway [38] out to 2100. First, we randomly draw 50,000 climate sensitivity samples from each distribution in Fig 1. With each climate sensitivity sample, we calculate the appropriate ocean diffusivity value from the calibration curve in Fig 2. We then project atmospheric temperature using the RCP8.5 radiative forcing time-series. This results in atmospheric temperature projected to year 2100 in an annual time-series for each climate sensitivity sample across all climate sensitivity distributions in Fig 1.

The second model we need maps the temperature projections to global sea-level rise (SLR). For this task, we use the SLR module described in the Regional Integrated Model of Climate and the Economy (RICE) [41]. This module maps changes in atmospheric temperature to changes in global sea level through contributions from thermal expansion, glacial melt, and ice sheets on Greenland and Antarctica. The atmospheric temperature output from the DOECLIM model runs described earlier are used as input into this module, producing a time-series of SLR out to year 2100 for each of the sampled climate sensitivities.

**Results and Discussion**

The distribution of climate sensitivity derived from the ensemble of CMIP models under-represent uncertainty in the SLR projections (Fig 4). The range of projected SLR in year 2100 produced by the model ensemble is 0.81 m – 1.12 m. This range misses between 8.5% and 91.8% of the mass of the distributions of projected SLR produced by the other climate sensitivity estimates which collectively range from 0.13 m to 1.25 m (Fig 5). While only a small portion of the mass that is missed when using the model ensemble is in the upper tail in this example, the difference can cause sizeable changes in risks and decisions about strategies to manage these risks by effectively shifting the failure rate of the designed infrastructure [34]. Additionally, tipping points, such as the collapse of portions of the Antarctic ice sheet, local atmosphere-ocean dynamics, and bathymetry may lead to even larger sea-level rise in some locations [42,43].

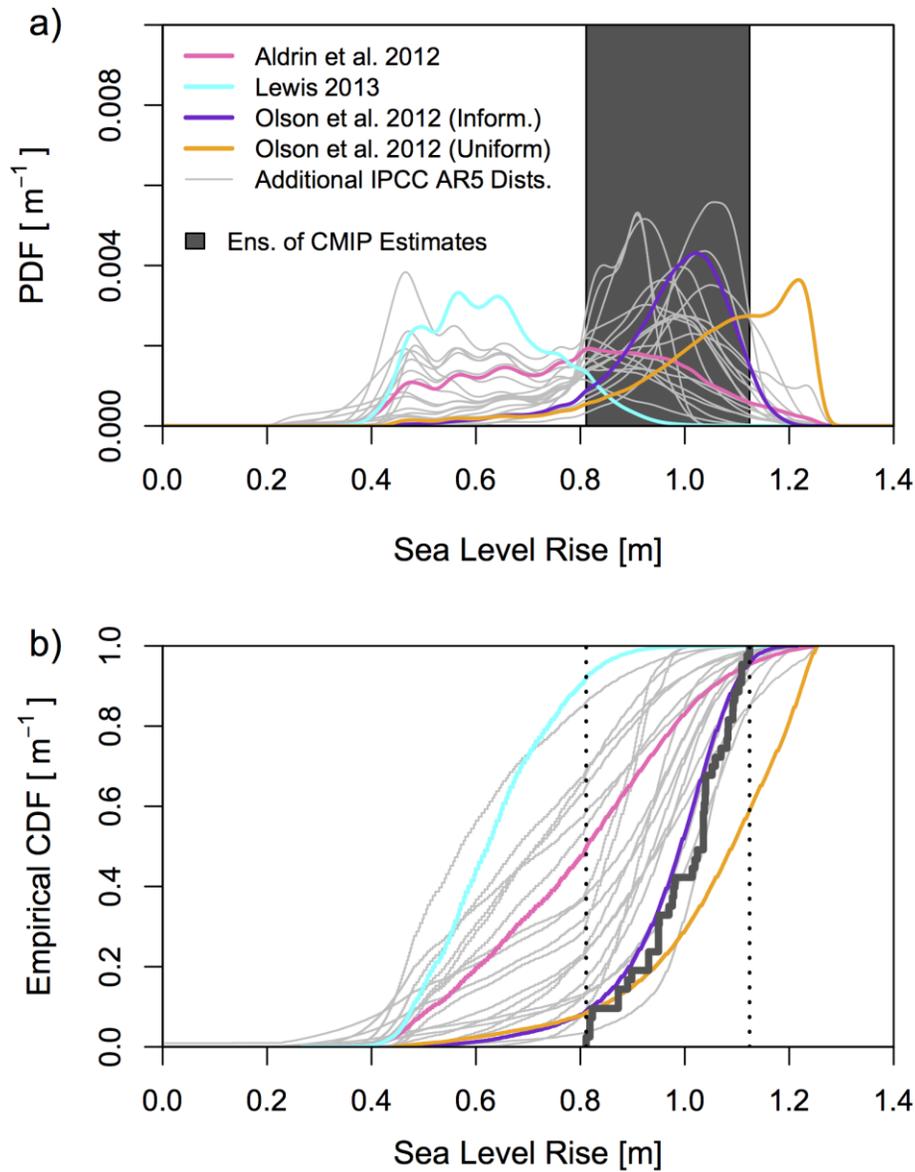

**Fig 4. Distribution of global mean SLR in projected year 2100.**

(A) The empirical probability distributions of SLR in projected year 2100 for each of the climate sensitivity distributions in Fig 1. The vertical black bar is the range of the CMIP model ensemble and the lines are the individual climate sensitivity distributions. The four distributions highlighted in color are consistent with the distributions highlighted in Fig 1. (B) The empirical cumulative distribution function (eCDF) of the SLR distributions in part A. Vertical dotted lines denote the range of SLR produced by the CMIP model ensemble.

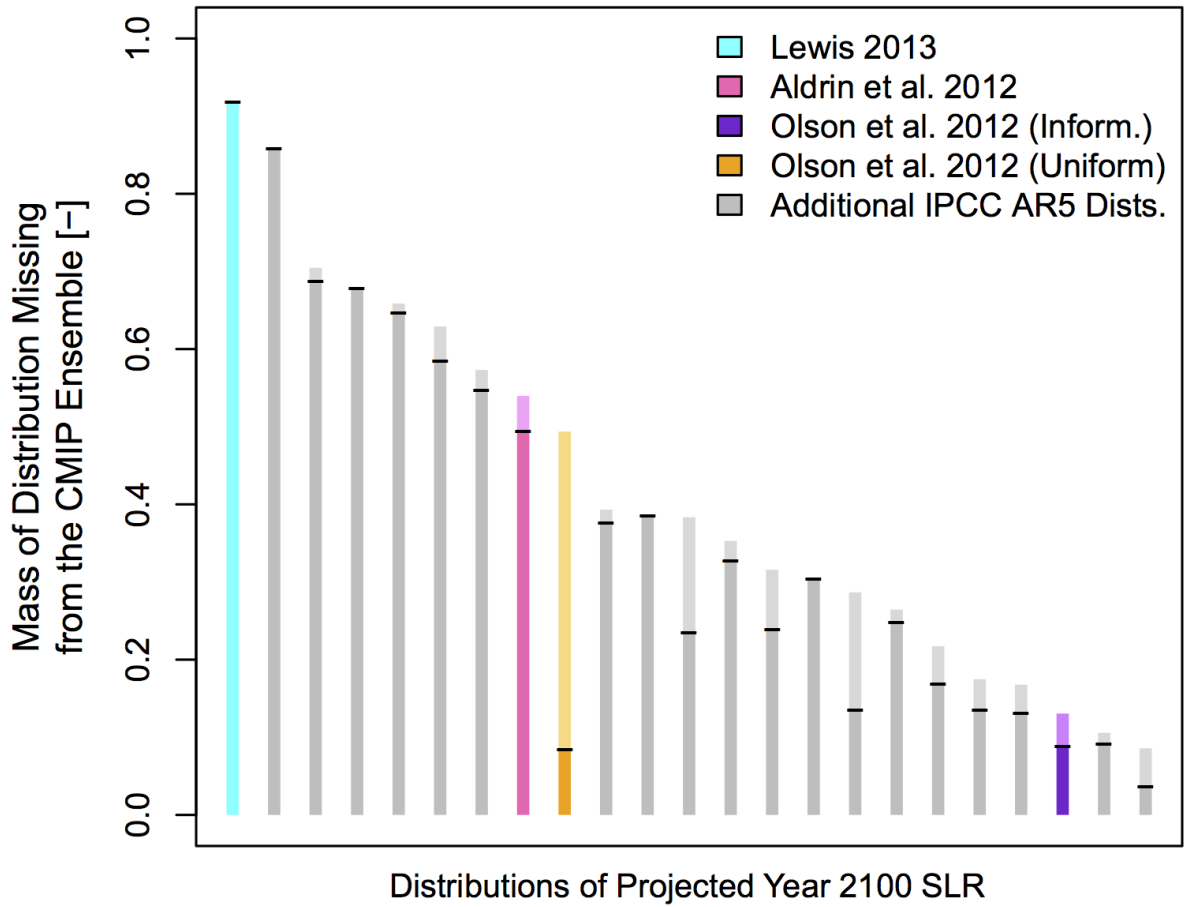

**Fig 5. CMIP model ensemble coverage of other climate sensitivity distributions.**

Vertical bars represent the integrated mass of the year 2100 SLR distributions that fall outside of the range of the CMIP model ensemble (i.e. the "cut off tails"). The horizontal black lines crossing each of the bars delimits the mass cut off from the lower tail (below – darker color) from the mass cut off from the upper tail (above – lighter color).

The effect of the tails of the distributions appears on decision-relevant timescales (Fig 6). Differences in the $0.5^{th}$ and $99.5^{th}$ percentiles between the model ensemble distribution and the other distributions become apparent as early as year 2030, which is 30 years into the simulation. This timescale is well within the considered lifetime of coastal flood defense infrastructures [1]. Considering more extreme percentiles of these distributions make these differences more apparent earlier in the simulation. These differences continue to increase throughout the duration of the simulation, suggesting that propagating the tails of these distributions becomes even more important in long-term planning and adaptation.

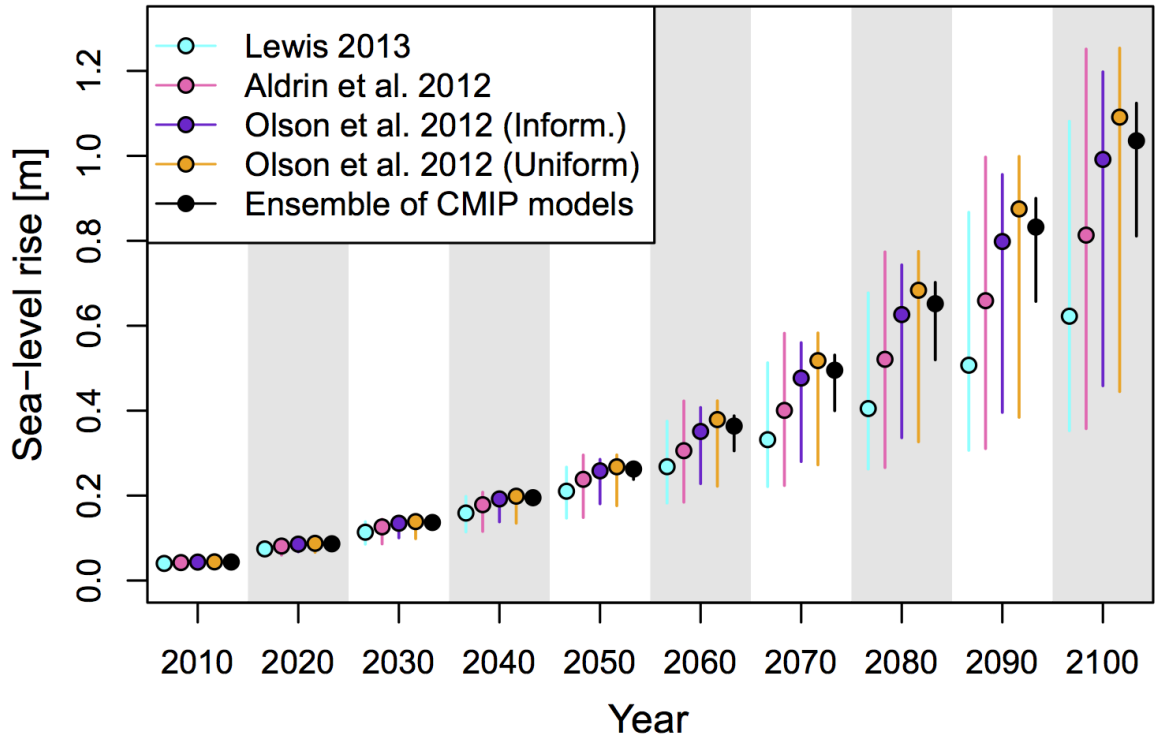

**Fig 6. Timeseries of projected SLR for the highlighted distributions and CMIP model ensemble.**

Dots represent the median and the vertical lines represent the 0.5$^{th}$ and 99.5$^{th}$ percentiles of the distributions of SLR for the appropriate simulation year. Only the highlighted distributions and the CMIP model ensemble are shown for clarity.

This analysis is designed to illustrate the problem of under-representing the tails of key climate components. To keep this illustration simple and transparent, the analysis is silent on important processes and uncertainties (for example changes in storm surge, land subsidence, or human behavior). The analysis shows, however, that the common approach of using an ensemble of complex-model outputs can under-represent low-probability potentially high-impact events that are decision relevant. Additionally, we show that reducing deep uncertainty to a single estimated distribution can affect decision-relevant metrics, thus leading the decision-maker to make decisions inconsistent with the real situation. This is especially important for decision analytical frameworks that are concerned about the robustness of a strategy over a broad range of possible future states of the world [1,23,24].

**Conclusions**

We strongly urge decision analysts to look beyond the status quo and consider a fuller representation of deep-uncertainty and decision-relevant tails in their analyses. Attention to decision-relevant features of the distributions of critical climate components better ensures that the analysis is both scientifically sound and ethically defensible. As our study demonstrates, this will require analysts to sample the known-unknowns that can influence these decision-relevant features by, for example, examining how uncertainty in the structure and parameter estimates of both the models and distributions affect the decision. This approach is more complex as it requires collaborations across academic disciplines (e.g. between decision analysts, statisticians, ethicists, Earth scientists, and engineers) as well as a meaningful engagement with real decision problems. However, an increased attention to decision-relevant features can help improve decisions and inform the design of mission-oriented basic science.

**Acknowledgments**

We thank A. Libardoni, C. Forest, D. Diaz, N. Tuana, P. Reed, R. Stouffer, M. Aldrin, R. Olson, and N. Urban for feedback on the manuscript. This work was partially supported by the National Science Foundation (NSF) through the Network for Sustainable Climate Risk Management (SCRiM) under NSF cooperative agreement GEO-1240507, NSF grant SES-1049208, and the Penn State Center for Climate Risk Management. Any opinions, findings, and conclusions or recommendations expressed in this material are those of the authors and do not necessarily reflect the views of the funding entities.

## References

1. Lempert R, Sriver R, Keller K. Characterizing Uncertain Sea Level Rise Projections To Support Investment Decisions. Calif Clim Chang Cent. 2012;

2. Van Dantzig D. Economic Decision Problems for Flood Prevention. Econometrica. 1956;24: 276–287.

3. Drouet L, Bosetti V, Tavoni M. Selection of climate policies under the uncertainties in the Fifth Assessment Report of the IPCC. Nat Clim Chang. 2015;5: 937–940. doi:10.1038/nclimate2721

4. Raisanen J, Palmer TN. A Probability and Decision-Model Analysis of a Multimodel Ensemble of Climate Change Simulations. J Clim. 2001;14: 3212–3226. doi:10.1175/1520-0442(2001)014<3212:APADMA>2.0.CO;2

5. Heyder U, Schaphoff S, Gerten D, Lucht W. Risk of severe climate change impact on the terrestrial biosphere. Environ Res Lett. 2011;6: 34036. doi:10.1088/1748-9326/6/3/034036

6. Rammig A, Jupp T, Thonicke K, Tietjen B, Heinke J, Ostberg S, et al. Estimating the risk of Amazonian forest dieback. New Phytol. 2010;187: 694–706. doi:10.1111/j.1469-8137.2010.03318.x

7. Hirabayashi Y, Mahendran R, Koirala S, Konoshima L, Yamazaki D, Watanabe S, et al. Global flood risk under climate change. Nat Clim Chang. 2013;3: 816–821. doi:10.1038/nclimate1911

8. Taylor KE, Stouffer RJ, Meehl GA. An overview of CMIP5 and the experiment design. Bull Am Meteorol Soc. 2012;93: 485–498. doi:10.1175/BAMS-D-11-00094.1

9. Tebaldi C, Knutti R. The use of the multi-model ensemble in probabilistic climate projections. Philos Trans R Soc A Math Phys Eng Sci. 2007;365: 2053–2075. doi:10.1098/rsta.2007.2076

10. O'Neill BC, Crutzen P, Grubler A, Ha-duong M, Keller K, Kolstad C, et al. Learning and climate change. Clim Policy. 2006;6: 585–589.

11. Keller K, Yohe G, Schlesinger M. Managing the risks of climate thresholds: uncertainties and information needs. Clim Change. 2008;91: 5–10. doi:10.1007/s10584-006-9114-6

12. National Research Council. Informing Decisions in a Changing Climate. Distribution. Washington, DC: The National Academies Press; 2009.

13. Hammitt JK, Shlyakhter AI. The Expected Value of Information and the Probability of Surprise. Risk Anal. 1999;19: 135–152.

14. Oppenheimer M, O'Neill BC, Webster M. Negative learning. Clim Change. 2008;89: 155–172. doi:10.1007/s10584-008-9405-1

15. Ricciuto DM, Davis KJ, Keller K. A Bayesian calibration of a simple carbon cycle model: The role of observations in estimating and reducing uncertainty. Global Biogeochem Cycles. 2008;22: 1–15. doi:10.1029/2006GB002908

16. Lempert R, Nakicenovic N, Sarewitz D, Schlesinger M. Characterizing climate-change uncertainties for decision-makers. An editorial essay. Clim Change. 2004;65: 1–9. doi:10.1023/B:CLIM.0000037561.75281.b3

17. Weitzman ML. Fat-Tailed Uncertainty in the Economics of Catastrophic Climate Change.

Rev Environ Econ Policy. 2011;5: 275–292. doi:10.1093/reep/rer006

18. Jonkman SN, Hillen MM, Nicholls RJ, Kanning W, van Ledden M. Costs of adapting coastal defences to sea-level rise— New estimates and their implications. J Coast Res. 2013;29: 1212–1226. doi:10.2112/JCOASTRES-D-12-00230.1

19. Steinschneider S, McCrary R, Mearns LO, Brown C. The effects of climate model similarity on probabilistic climate projections and the implications for local, risk-based adaptation planning. Geophys Res Lett. 2015;42: 5014–5022. doi:10.1002/2015GL064529

20. Sexton DMH, Murphy JM. Multivariate probabilistic projections using imperfect climate models. Part II: robustness of methodological choices and consequences for climate sensitivity. Clim Dyn. 2012;38: 2543–2558. doi:10.1007/s00382-011-1209-8

21. Katz RW, Craigmile PF, Guttorp P, Haran M, Sansó B, Stein ML. Uncertainty analysis in climate change assessments. Nat Clim Chang. 2013;3: 769–771. doi:10.1038/nclimate1980

22. Lempert RJ, Schlesinger ME. Robust Strategies for Abating Climate Change. Clim Change. 2000;45: 387–401. doi:10.1023/A:1005698407365

23. Weaver CP, Lempert RJ, Brown C, Hall JA, Revell D, Sarewitz D. Improving the contribution of climate model information to decision making: The value and demands of robust decision frameworks. Wiley Interdiscip Rev Clim Chang. 2013;4: 39–60. doi:10.1002/wcc.202

24. Herman JD, Reed PM, Zeff HB, Characklis GW. How Should Robustness Be Defined for Water Systems Planning under Change? J Water Resour Plan Manag. 2015;141: 4015012. doi:10.1061/(ASCE)WR.1943-5452.0000509

25. Sherwood SC, Bony S, Dufresne J-L. Spread in model climate sensitivity traced to atmospheric convective mixing. Nature. 2014;505: 37–42. doi:10.1038/nature12829

26. Hansen J, Lacis A, Rind D, Russel G, Stone P, Fung I, et al. Climate Sensitivity: Analysis of Feedback Mechanisms. Climate Processes and Climate Sensitivity. American Geophysical Union; 1984. pp. 130–163. doi:10.1029/GM029p0130

27. Aldrin M, Holden M, Guttorp P, Skeie RB, Myhre G, Berntsen TK. Bayesian estimation of climate sensitivity based on a simple climate model fitted to observations of hemispheric temperatures and global ocean heat content. Environmetrics. 2012;23: 253–271. doi:10.1002/env.2140

28. Olson R, Sriver R, Goes M, Urban NM, Matthews HD, Haran M, et al. A climate sensitivity estimate using Bayesian fusion of instrumental observations and an Earth System model. J Geophys Res Atmos. 2012;117: 1–11. doi:10.1029/2011JD016620

29. Lewis N. An Objective Bayesian Improved Approach for Applying Optimal Fingerprint Techniques to Estimate Climate Sensitivity. J Clim. 2013;26: 7414–7429. doi:10.1175/JCLI-D-12-00473.1

30. Bindoff NL, Stott PA, AchutaRao KM, Allen MR, Gillett N, Gutzler D, et al. Chapter 10 - Detection and attribution of climate change: From global to regional. Climate Change 2013 - The Physical Science Basis IPCC Working Group I Contribution to AR5. Cambridge University Press; 2013. Available: http://pure.iiasa.ac.at/10552/

31. Stainforth DA, Aina T, Christensen C, Collins M, Faull N, Frame DJ, et al. Uncertainty in predictions of the climate response to rising levels of greenhouse gases. Nature. 2005;433: 403–406. doi:10.1038/nature03301

32. Budescu D V., Broomell SB, Lempert RJ, Keller K. Aided and unaided decisions with imprecise probabilities in the domain of losses. EURO J Decis Process. 2014;2: 31–62. doi:10.1007/s40070-013-0023-4

33. Keller K, Nicholas R. Improving climate projectinos to better inform climate risk management. In: Bernard L, Semmler W, editors. The Oxford Handbook of the Macroeconomics of Global Warming. Oxford: Oxford University Press; 2014.

34. Sriver RL, Urban NM, Olson R, Keller K. Toward a physically plausible upper bound of sea-level rise projections. Clim Change. 2012;115: 893–902. doi:10.1007/s10584-012-0610-6

35. Kirtman B, Power SB, Adeodoyin AJ, Boer GJ, Bojariu R, Camilloni I, et al. Near-term Climate Change: Projections and Predictability. In: Stocker TF, Qin D, Plattner G-K, Tignor M, Allen SK, Boschung J, et al., editors. Climate Change 2013: The Physical Science Basis Contribution of Working Group I to the Fifth Assessment Report of the Intergovernmental Panel on Climate Change. Cambridge, United Kingdom and New York, NY, USA: Cambridge University Press; 2013. Available: https://www.ipcc.ch/pdf/assessment-report/ar5/wg1/WG1AR5_Chapter11_FINAL.pdf

36. Kriegler E. Imprecise Probability Analysis for Integrated Assessment of Climate Change [Internet]. Universitat Potsdam. 2005. Available: https://www.pik-potsdam.de/members/edenh/theses/PhDKriegler.pdf

37. Goes M, Tuana N, Keller K. The economics (or lack thereof) of aerosol geoengineering. Clim Change. 2011;109: 719–744. doi:10.1007/s10584-010-9961-z

38. Meinshausen M, Smith SJ, Calvin K, Daniel JS, Kainum MLT, Lamarque J-F, et al. The

RCP greenhouse gas concentrations and their extensions from 1765 to 2300. Clim Change. 2011;109: 213–241. doi:10.1007/s10584-011-0156-z

39. Hansen J, Ruedy R, Sato M, Lo K. Global surface temperature change. Rev Geophys. 2010;48: RG4004. doi:10.1029/2010RG000345

40. GISS Team. GISS Surface Temperature Analysis (GISTEMP) [Internet]. 2017 NASA Goddard Institute for Space Studies [cited 25 Oct 2017]. Available: https://data.giss.nasa.gov/gistemp/

41. Nordhaus WD. Economic aspects of global warming in a post-Copenhagen environment. Proc Natl Acad Sci U S A. National Academy of Sciences; 2010;107: 11721–6. doi:10.1073/pnas.1005985107

42. Kopp RE, Hay CC, Little CM, Mitrovica JX. Geographic Variability of Sea-Level Change. Curr Clim Chang Reports. Springer International Publishing; 2015;1: 192–204. doi:10.1007/s40641-015-0015-5

43. DeConto RM, Pollard D. Contribution of Antarctica to past and future sea-level rise. Nature. 2016;531: 591–597. doi:10.1038/nature17145